\documentstyle[12pt,epsf]{article}
\begin{document}
\begin{center}
{\bf Temperature, Pressure \& Solvent Dependence of 
Positronium Acceptor Reactions}
\end{center}
\begin{center}
{\sl Debarshi Gangopadhyay$^*$, Bichitra Ganguly$^*$ 
\& Binayak Dutta-Roy$^\dagger$}
\end{center}
\begin{center} 
$^*$ Saha Institute of Nuclear Physics, 
I/AF Bidhannagar, Calcutta 700 064, India
\end{center}
\begin{center}
 $^\dagger$ S.N. Bose National Centre for Basic Sciences,
III/JD Bidhannagar, Calcutta 700 098
\end{center}
\begin{center}
\end{center}
\vskip .1truecm
\begin{abstract}
Positronium (Ps) reaction rates ($\kappa$) with weak Acceptors (Ac) leading to 
the formation of Ps-Ac Complexes show 
several intriguing features: non-monotonic temperature dependence of $\kappa$ 
(departing from the usual Arrhenius paradigm), considerable variability of 
$\kappa$  
with respect to different solvents, and anomalies in response to external 
pressure at ambient temperature (large changes of 
$\kappa$ in some media and hardly 
any in others). We explain all these
phenomena, introducing the novel concept of a critical 
surface tension, which unifies observations in diverse 
non-polar solvents at different temperatures and pressures.
\end{abstract}
\vskip .1truecm
The Positronium (Ps) atom, as the `lightest isotope of hydrogen', enters 
into various chemical reactions, amongst which we shall concentrate here 
on those with diamagnetic organic compounds (equipped with a suitable 
low-lying molecular orbital) such that a Positronium-Acceptor (Ps-Ac) 
Complex can form in the presence of some solvent (S).

The observed rate constant ($\kappa$) for a given Acceptor depends on the 
solvent and for a given solvent varies in a rather remarkable manner with 
temperature ($T$), namely:

\noindent $\bullet$ $\kappa$ increases with $T$ at low temperatures,

\noindent $\bullet$ $\kappa$ reaches a maximum at $T=T_0$ (the turnover 
temperature),

\noindent $\bullet$ $\kappa$ decreases with increasing $T$ above $T_0$.

\noindent This is shown in Fig.1 for the case of the weak Acceptor 
nitrobenzene in different solvents.

\begin{figure}[htbp]
\epsfxsize=7.5cm
\epsfysize=8cm
\centerline{\epsfbox{fig1.prn}}
Fig.1 \it{ Observed temperature dependence of rate constants for Ps reaction
with nitrobenzene in different solvents. (Taken from W.D.Madia et.al.,
J.Am.Chem.Soc. 97(1975)5041). The lines correspond to slopes in the enthalpy
dominated and diffusion controlled regimes as predicted by our model(with
appropriate over-all normalization for heptane).}
\end{figure}

 This unusual phenomenon [1], first
observed by Goldanskii and his group [2] at Moscow, and subsequently 
by Hall, Madia and Ache [3], is in sharp contrast with the normal trend 
in activated chemical processes where $ln~\kappa$ versus $1/T$ (the 
inverse temperature) yields a straight line with negative slope, as the 
rate is proportional to the Maxwell-Boltzmann probability factor $\exp 
(-E^*/k_B T)$ where $k_B$ is the Boltzmann constant and $E^*$ is the 
activation energy or barrier height to be overcome by thermal agitation. 
Possible prefactors are generally taken to be mildly dependent on 
temperature.

Another dramatic observation [4,5] pertains to the variation of the 
rate constant for Ps-Ac Complex formation with external pressure at 
ambient temperature, which shows surprisingly strong solvent dependence.
Thus, for instance, with nitrobenzene as the Acceptor and hexane as the 
solvent the rate constant $\kappa$ was found to be enhanced by a factor 
of almost thirty as the pressure was increased to $\sim$ 1000 kg/cm$^2$, 
while with benzene as the solvent the same rate manifested only small 
variation over the same pressure range. This feature is depicted in 
Fig.2. Kobayashi [4,5] conjectured that this could indicate a 
difference in reaction mechanism in cases which do and those which do not 
show marked pressure dependence, and that a correlation could be sought 
with the corresponding solvent dependence of the turnover temperature.
\begin{figure}[htbp]
\epsfxsize=7.5cm
\epsfysize=8cm
\centerline{\epsfbox{fig2.prn}}
Fig.2 {\it Observed dependence on external pressure for Ps reaction with
nitrobenzene in hexane and in benzene. The predicted dependence in the case
of hexane is shown by the solid line (appropriate normalization has been done).} 
\end{figure}
\vskip .2cm
The objective of this letter is to put forward a heuristic model which 
captures the essence of all these puzzling aspects of Positronium reactions 
outlined above, through a remarkably simple and reasonably accurate 
semi-quantitative  description.
As we shall see, the widely used bubble model for Ps-annihilation in liquids 
provides a suitable setting for the purpose at hand. This was proposed by 
Ferrel [6] to account for the observed largeness of the ortho-Positronium 
(o-Ps) pick-off lifetime, whereby the positron in o-Ps seeks out an electron 
in the surrounding medium with opposite spin (not its partner in o-Ps which
is in a spin triplet state) to 
decay into two gammas. He suggested that due to the repulsive interaction 
between o-Ps and the atoms of the surrounding liquid, the Positronium 
pushes away the molecules of the liquid and gets self-trapped in a cavity 
or bubble. The decreased probability for the positron to find an electron 
(of the surrounding medium) in its immediate vicinity leads to the 
lengthening of the pick-off lifetime. For simplicity he took the trapping 
potential to be an infinitely repulsive spherical well of radius $R$. 
The resulting quantal zero-point energy of Ps with mass $2m$ ( $m$ being 
the electron mass) is 
$
E_0 = {\pi^2 \hbar^2 \over 4m R^2}.
$
This exerts an outward `pressure' viz. $-{\partial E_0 \over \partial R}$ 
to be balanced by the contractile forces of compression due to the 
surface tension ($\sigma$) of the fluid, thereby minimizing the total energy 
$E_{tot} = E_0 + 4 \pi R^2 \sigma$ viz. ${\partial E_{tot} \over \partial R} 
=-{2\pi^2\hbar^2\over 4m R^3} + 8\pi \sigma = 0$ resulting in a bubble of 
radius
$
R_0 = ({\pi\hbar^2 \over 16 m \sigma})^{1/4} = (12.445)/\sigma^{1/4} 
$\AA$~~$,
where 
 $\sigma$ is measured in dynes/cm ( .624$\times$ 10 $^{-4}$ eV/\AA$^2$).
This model was further developed by Tao [7] and by Eldrup et al. [8] employing 
again the infinite spherical well but describing the picked-off electrons as 
forming a thin layer uniformly coating the inner surface of the bubble. 
Though such a description has been criticised on account of the infinite 
repulsion [9,10] or the unrealistic nature of the sharpness of the bubble 
edge [11,12], nevertheless as a first approximation the model does rather 
well in explaining the observables in the case of pure solvents (lifetime 
and angular correlation of decay gammas).

In this framework the formation of the Ps-Ac Complex within the cavity 
results, through the restriction on the motion of the Positronium due to
its binding to the Acceptor, in a release of the outward `uncertainty pressure' and accordingly the bubble begins to shrink under the influence of the 
surface tension. As the bubble wall approaches the Positronium (bound to 
the Acceptor with some energy $E=-B$ where $B$ is the binding energy), the 
increasing proximity of the solvent molecules leads (because of the augmented 
repulsive energy) to a reduction of the binding, until at some radius $R_b$ 
of the bubble the Ps-Ac bond is broken viz. $B=0$. To raise the status of 
this contention [13] to at least a semi-quantitative level let us 
consider the Positronium to be subjected to an attractive potential $V(r)$ 
due to the Acceptor, while it also feels the infinite repulsion due to the 
solvent located at the bubble wall at a distance $R$. Taking the potential 
$V(r)$ to be a spherical well [14] of depth $V_0$ and range $a$ viz. 
$-V_0 \Theta (a-R)$ where $\Theta$ is the step function, the relevant 
Schr\"odinger equation for the centre-of-mass motion of Ps is 
$$
\Bigl[-{\hbar^2\over 4m}{d^2\over dr^2} -V_0\Bigr]u = -B u~~~{\rm for}~~~r < a
\eqno{(1a)}
$$
$$
\Bigl[-{\hbar^2\over 4m}{d^2\over dr^2}\Bigr]u=-B u ~~~{\rm for}~~~r > a
\eqno{(1b)}
$$
with $\psi = u/r$. The appropriate boundary conditions are $u(r=0)=0$ 
and $u(r=R)=0$, while $u$ and its derivative
must be continuous at $r=a$, yielding the eigenvalue condition 
$$
k\tan \kappa a = -\kappa \tanh k(R-a)
\eqno{(2)}
$$
where $k={\sqrt {4m B\over \hbar^2}}$ and $\kappa ={\sqrt{{\tilde g}^2 
-k^2}}$ with ${4m V_0\over \hbar^2}={\tilde g}^2$.
The bond breaking radius $R_b$, viz. where $k=0$, is thus given by 
$$
R_b = a \Bigl[1-{\tan{\tilde g}a\over {\tilde g}a}\Bigr].
\eqno{(3)}
$$
Since in the leading approximation (adopted by us) the effect of 
the solvent is represented by an infinite repulsion (tantamount to a 
boundary condition forcing $\psi$ to vanish at $R$), the radius $R_b$ 
depends only on the characteristics of the Acceptor vis a vis its affinity 
for the Positronium, and is, to that extent, same for all solvents.
However, we must also consider the fact that the Ps-Ac-bubble system 
would equilibrate at some radius $R_{min}$ at which the total energy is 
a minimum, namely,
$$
{\partial \over \partial R} [-B + 4\pi R^2 \sigma] = 0.
\eqno{(4)}
$$
This leads to 
$$
R_{min} = ({1\over 8\pi\sigma}) ({\hbar^2\over 4m})({1\over a^4}) 
{6\zeta^4 \over (\zeta -\tan \zeta)(3\zeta -3 \tan\zeta +3\zeta \tan^2 \zeta 
-2\tan^3\zeta)}
\eqno{(5)}
$$ 
where $\zeta \equiv {\tilde g} a$. Whether the Ps-Ac-bubble system will be 
stable or not will depend on whether $R_{min}$ is greater than or lesser 
than the bond-breaking radius $R_b$.

A big step forward is made possible by our recognition that in leading 
approximation $R_b$ is essentially solvent independent, while solvent 
dependence enters through $R_{min}$ and that too only through $\sigma$. 
As the surface tension of the solvent is a function of temperature, the 
equilibrium radius of the bubble (with Ps-Ac inside) depends on temperature 
only through $\sigma$ viz. $R_{min} (T) = R_{min} [\sigma (T)]$. As $T$ is 
decreased, $\sigma$ increases and by virtue of eq.(4) and more explicitly 
eq.(5), $R_{min}$ decreases. When $R_{min}$ becomes less than $R_b$ the 
Ps-Ac-bubble system is no longer stable. Clearly there is a critical value 
of $\sigma$, say $\sigma_{cr}$, at which $R_{min}[\sigma_{cr}] = R_b$, 
marking the watershed beyond which the Ps-Ac-bubble system destabilises. 
The value of $T$ at which $\sigma$ attains the value $\sigma_{cr}$ depends 
on the particular solvent, but $\sigma_{cr}$ does not, but is, in the leading 
order, a property of the Acceptor under consideration. We argue that the 
turnover temperature $T_0$ is the temperature at which, for the solvent at hand, 
$\sigma (T_0) = \sigma_{cr}$. Indeed at sufficiently high temperature 
where $\sigma (T) < \sigma_{cr}$, a large negative change in enthalpy 
occurs as the reactants [Ps in a bubble of radius $R_0 = ({\pi \hbar^2 \over 
16 m \sigma})^{1/4}$ and the Acceptor] react to form the product 
[Ps-Ac Complex in bubble] with the radius having shrunk to $R_{min}$. This 
makes it a down-hill reaction, since for weak Acceptors the activation energy 
is expected to play a sub-dominant role and also the effect of solvent 
viscosity is negligible (because the complex is 
protected by the bubble from the 
buffetting by the solvent molecules). It is thus basically the negative 
activation volume that is responsible for the anti-Arrhenius behaviour for 
$T > T_0$. On the other hand for $T < T_0$ when $\sigma (T) > \sigma_{cr}$ 
the Ps-Ac-bubble system is unstable, the Ac is squeezed out of the bubble 
and the formation of the Complex must of necessity take place in the milieu 
of the solvent accompanied by the continual impact of the solvent molecules. 
As such, following Kramers [15] the role of the medium would be said to 
belong to the Smolochowski regime, with the dependence of the rate constant 
on the viscosity ($\eta$) varying as ${1\over \eta} \sim D$ (where $D$ is 
the diffusion coefficient of the liquid, the last step being a consequence 
of the Einstein-Stokes relation). In view of the smallness of the activation 
energy for weak Acceptors, we would therefore expect that $\kappa \sim 
\exp[-E_\eta/k_B T]$, where $E_\eta$ is the activation energy associated with 
the process of diffusion. The reaction in this region would be diffusion 
controlled and would exhibit a normal Arrhenius behaviour. This enables us 
to make the important prediction that $\sigma (T_0) = \sigma_{cr}$ and that 
it is approximately solvent independent. In order to confront this with 
experiment, we have plotted in Fig.3, $\sigma (T)$ against $T-T_0$ for 
various solvents (for which data is available with nitrobenzene as the 
Acceptor).

\begin{figure}[htbp]
\epsfxsize=8cm
\epsfysize=8cm
\centerline{\epsfbox{fig3.prn}}
Fig.3 {\it Surface tension $\sigma(T)$ for different solvents as a function of
$T-T_0$ exposing the concept of the critical surface tension ($\sigma_{cr}$)}
\end{figure}

 It is indeed highly gratifying to note that while $T_0$ differs
widely from solvent to solvent, and also the values of the surface tension 
$\sigma$ at a given temperature for different solvents have a 
substantial spread, nevertheless $\sigma (T_0)=\sigma_{cr}$ for the 
solvents under consideration lie in a rather narrow range, namely, 
$\sigma_{cr} = 26 \pm 2$ dynes/cm. 

Armed with this value of $\sigma_{cr}$ we are now in a position to access the 
Ps-Ac interaction parameters which, as we shall discuss below, 
are unfortunately 
not available from any other source. Being led by the estimates of other 
authors [16] let us fix the range of the interaction to be $a$ = 1.5 \AA$~$, 
and determine $\tilde g$ by putting, in accordance to our discussions above, 
$\sigma = \sigma_{cr}$ and $R_{min} = R_b$ [refer eq.(3)]. This immediately 
yields the value ${\tilde g}$ =1.25 \AA$^{-1}$ which corresponds to the 
Ps-Ac binding energy (the Acceptor being nitrobenzene) $B_0 = 0.18 eV$. Note 
that $B_0$ signifies the basic binding in the absence of the solvent, namely, 
$B_0 = Lim_{R \rightarrow \infty} B$, which implies vide eq.(2), the 
well-known result:
$
-\kappa \cot \kappa a = k
$
the eigenvalue condition for a spherical well. Unfortunately there is 
no direct measurement of this binding energy $B_0$, and the approaches  
based on first principles [17] are beset with huge theoretical uncertainties. 
However, using our methodology we have now completely tied down the model 
and shall proceed to show how this explains all the main observed features,
   
In order to find the slope of the $ln~ \kappa$ versus $1/T$ plot in the 
higher temperature anti-Arrhenius region we recognise that this is nothing 
but the activation free energy, the major contribution to which come from 
the change in Enthalpy arising from the shrinkage of the bubble from its 
value given by $R_0$ to the size determined by $R_{min}$, viz.
$
4 \pi R_{min}^2 \sigma - 4 \pi R_0^2 \sigma
$
where, of course the radii are functions of temperature through the 
surface tension. To this must be added the change in the Ps-Ac binding 
due to the approach of the bubble wall from $R_0$ to $R_{min}$. The 
resulting behaviour (appropriately normalized) is depicted in Fig.1 for the
case of Ps nitrobenzene
reaction in heptane, for instance, as a solvent. The slope obtained through our model 
corresponds to $\approx$ 0.15 eV which agrees very well with the experimental 
value.
 The part of the plot in the Arrhenius 
regime is also shown by using the value of the activation energy for 
diffusion ($E_\eta$). Fig.1 demonstrates that 
the main aspect of the temperature dependence of the rate has been 
captured in a satisfactory manner.

Apart from providing an interpretation of the temperature dependence of the 
reaction as set forth above, this simple model is also able to furnish an 
explanation of the observed variation with the external pressure (P). In 
view of the small compressibility of liquids one would hardly expect any 
appreciable activation volume when $T < T_0$ as the reaction occurs in 
the solvent itself. However, for $T > T_0$ due to the involvement of the 
bubble a large $\Delta V^*$ (activation volume) becomes possible and hence 
a significant pressure dependence can occur, in view of the basic Polyani 
relationship:
$$
\Biggl( {\partial ln \kappa \over \partial P}\Biggr)_T = -{\Delta V^*\over k_B T}
\eqno{(6)}
$$
between the effect of external pressure on the rate of a chemical reaction 
and the activation volume of the reaction (namely, the difference between 
the volume of the activated complex and the volume of the reactants). To 
obtain quantitative estimates it is to be noted that the initial volume viz. 
Ps in the bubble without the Ac must be found now by minimizing the total 
energy, namely
$
E_{tot} = {\hbar^2\over 4mR^2} + 4\pi R^2 \sigma + {4\pi\over 3} R^3 P 
$
with respect to $R$ and thereby obtain $R_0(P)$. Here $P$ is measured in 
kgwt/cm$^2$ (=$ .613 \times 10^{-6}$ eV/\AA$^3$). Similarly the equation 
determining the equilibrium radius [$R_{min}(P)$] with Ps-Ac inside the bubble 
must be appropriately modified, so that in place of eq.(4), we now have
$$
{\partial\over \partial R} \Bigl( -B + 4\pi R^2 \sigma + {4\pi\over 3}
 R^3 P\Bigr) = 0
\eqno{(7)}
$$
Equipped with $R_0(P)$, $R_{min}(P)$ and $B[R(P)]$ the variation of the rate 
constant with pressure can be readily calculated. 

Consider the two typical cases depicted in Fig.2 out of the many taken 
from the experimental studies of Kobayashi [4,5]. All these experiments were 
performed at ambient temperature ($T_{expt} = 19\pm 1^oC$). With nitrobenzene 
as the Acceptor and benzene as the solvent we note that there is no significant
effect of pressure on the reaction rate. This is quite consistent with the model
in view of the fact that the turnover temperature in this case [18] is
considerably above $T_{expt}$ and hence one is working in the Arrhenius region 
where the pressure effect is expected to be small. On the other hand when 
the solvent is hexane, the experiment was performed in the regime where the 
Ps-Ac-bubble system is stable (as the turnover temperature here is -53$^oC$), 
and accordingly the reaction rate responds appreciably to external pressure.
Indeed the enhancement of the rate by a factor of about thirty in our 
model is even in quantitative agreement with the experimental results of 
Kobayashi. 
Using the experimental value of the surface tension of hexane at $T_{expt}$ 
($\sigma\approx$ 19 dynes/cm), the model indicates the behaviour shown in 
Fig.2 with appropriate over-all normalization.

 Above a critical pressure the Ps-Ac-bubble system destabilizes
and the rate versus pressure curve levels off. Except for the fact that there 
is a somewhat precocious onset of the Arrhenius regime the general experimental 
trend is captured very well indeed considering the approximations involved.

We may thus conclude that the simple model we have proposed effectively 
accounts for the temperature, pressure and solvent dependence of Ps-Ac 
Complex formation reactions. Of cardinal importance for our discussion 
has been the introduction of the notion of the critical surface tension 
($\sigma_{cr}$) which enables us to semiquantitatively account for most 
observations and also to make the important qualitative remark: 
that for any solvent at a temperature such that its 
surface tension is smaller than $\sigma_{cr}$ the reaction rate decreases 
with increasing temperature and is also significantly affected by external 
pressure, while otherwise it is the other way around. 
\vskip .3truecm

\vskip .2truecm
\noindent [1] O.E. Mogensen, Positronium Annihilation in Chemistry (Springer-
Verlag, 1995).
\\
\noindent [2] V.I. Goldanskii {\it et al.}, 
Dokl. Acad. Nauk. {\bf 203} (1972) 870. 
\\
\noindent [3] E. Hall, W.J. Madia \& H.J. Ache, Radiochem. Radioanal. Lett. 
{\bf 23} (1975) 283.
\\
\noindent [4] Y. Kobayashi, Chem. Phys. Lett. {\bf 172} (1990) 307.
\\
\noindent [5] Y. Kobayashi, Chem. Phys. {\bf 150} (1991) 453.
\\
\noindent [6] R.A. Ferrel, Phys. Rev. {\bf 16} (1957) 108.
\\
\noindent [7] S.J. Tao, J. Chem. Phys. {\bf 56} (1972) 5499.
\\
\noindent [8] M. Eldrup, D. Lightbody \& J.N. Sherwood, Chem. Phys. {\bf 63} 
(1981) 51.
\\
\noindent [9] L.O. Roellig,  Positron Annihilation, ed. 
A.T. Stewart \& L.O. Roellig (Academic Press, N.Y.) p 127.
\\
\noindent [10] T.B. Daniel \& R. Stump, Phys. Rev. {\bf 115} (1959) 1599.
\\
\noindent [11] T. Mukherjee, B. Ganguly \& B. Dutta-Roy, J. Chem. Phys. 
{\bf 107} (1997) 7467.
\\
\noindent [12] D. Gangopadhyay {\it et al.}, 
J. Phys. Cond. Matt. {\bf 11} (1999) 1463.
\\
\noindent [13] A similar two potential model was suggested in A.P. Buchikin, 
V.I. Goldanskii \& V.P. Shantarovich, Dokl. Acad. Nauk. {\bf 212} (1973) 
1356 and in A.I. Ryzhkov \& V.P. Shantarovich, Mat. Sci. Forum {\bf 105-108} 
(1992) 1699. However, with both potentials having finite depths there were 
too many parameters to give the model predictive power or provide a deeper 
insight.
\\
\noindent [14] The detailed form of the binding potential hardly matters. 
We have checked that with a Dirac delta potential $V(r) = -g \delta (r-a)$, 
the results are very similar. It must however be mentioned that even though 
the general trends in each of these models essentially remain the same 
detailed values are often very sensitively dependent on the numerical values 
of parameters.
\\
\noindent [15] H.A. Kramers, Physica {\bf 7} (1940) 284.
\\
\noindent [16] V.I. Goldanskii \& V.P. Shantarovich, Appl. Phys. {\bf 3} (1974) 
335. The range 1.5 \AA $~$ represents the distance
between the active centres i.e. electron of Ps and that of C$^+$ ion.
\\
\noindent [17] D.M. Schrader \& C.M. Wang, J. Phys. Chem. {\bf 80} (1976) 2507. 
\\
\noindent [18] Indications are already there in reference 2 but very careful 
measurements recently made in our laboratory place the turnover temperature 
for Ps-nitrobenzene reaction with benzene as a solvent at $33\pm 3^oC$.  

\end{document}